\documentclass[fleqn,usenatbib]{mnras}

\usepackage{graphicx} % For figures
\usepackage{amsmath} % For advanced math
\usepackage[T1]{fontenc} % Font encoding
\usepackage{txfonts} % Consistent with your original
\usepackage{color} % For colored text (if needed)
\newcommand*{\mk}{\color{black}} % Your custom command

\title[Magnetic, Kinetic, and Transition Regimes]{Magnetic, Kinetic, and Transition Regime: Spatially-Segregated Structure of Compressive MHD Turbulence}
\author[Li et al.]{Guang-Xing Li,$^{1}$ Mengke Zhao$^{1,2}$\\
$^{1}$South-Western Institute for Astronomy Research, Yunnan University, Chenggong District, Kunming 650500, China\\
$^{2}$School of Astronomy and Space Science, Nanjing University, 163 Xianlin Avenue, Nanjing 210023, People’s Republic of China\\
Correspondence: gxli@ynu.edu.cn}

\begin{document}
\maketitle

\begin{abstract}
Turbulence is a complex physical process prevalent in modern physics, particularly in ionized environments like interstellar gas, where magnetic fields play a dynamic role. However, the precise influence of magnetic fields in such settings remains unclear. We employ the Alfvén Mach number, ${\cal M}_{\rm A} = \sqrt{E_{\rm k}/E_B}$, to gauge the magnetic field’s significance relative to turbulent motion, uncovering diverse interaction patterns. In the low-${\cal M}_{\rm A}$ magnetic regime, the field is force-free, yet gas motion does not align with it. At intermediate ${\cal M}_{\rm A}$ (magnetic-kinetic transition regime), velocity and magnetic fields show peak alignment, likely due to rapid relaxation. In the high-${\cal M}_{\rm A}$ kinetic regime, both fields are irregular and unaligned. These regimes find observational counterparts in interstellar gas, highlighting the multifaceted nature of MHD turbulence and aiding future astrophysical interpretations.
\end{abstract}

\begin{keywords}
Magnetic fields -- Magnetohydrodynamics (MHD) -- ISM: clouds -- ISM: kinematics and dynamics -- ISM: magnetic fields
\end{keywords}
% \section{Introduction}

\section{Introduction}
% MHD turbulence, Role of magnetic field. Ma

Turbulence is a complex process that has puzzled scientists since the age of Leonardo da Vinci, and the capability of turbulence in controlling the evolution of interstellar gas has been known for decades 
\citep{2004RvMP...76..125M}.
Understanding the role of the magnetic field in compressible turbulence can be crucial for our understanding of turbulence and for interpreting astrophysical observations. Turbulence is a complex, multi-scale process  \citep{1995tlan.book.....F} best described using scaling relations \citep{1941DoSSR..30..301K}. Past studies of compressible magnetohydrodynamics (MHD) turbulence have followed this tradition where describing the statistic properties of the region has become a priority \citep{2020ApJ...905...14B}. Others have astrophysical applications in mind and have focused on the global quantities extracted from the simulation box \citep{2011ApJ...730...40P}.

The alignment between vector quantities such as the magnetic field $\vec{B}$, the velocity $\vec{v}$, and the current $J$ offers insight into the behavior of the magnetized fluids. In astrophysical research, the alignment between $\vec{B}$ and $\vec{v}$ is assumed to be an indicator of the magnetic field's ability to affect the motion of the gas. The alignment between $\vec{B}$ and $\vec{v}$ in the strongly magnetized
 regime is a fact often taken for granted, and this picture is the foundation for understanding phenomena such as wind from disk-star systems \citep{2007prpl.conf..277P}, where the picture of \emph{beads on a wire} have been widely accepted. In this picture, field lines of a magnetic field behave like rigid wires, which guide the motion of the gas.
 However, it is unclear to what extent can we trust this picture. On the other
 hand, alignment between $\vec{B}$ and $\vec{v}$ has been proposed by
 \citep{2006PhRvL..96k5002B}. However, the alignment is analyzed in the Fourier
 space, where the spatial structure and possible segregation effect can not be
 analyzed.
 \cite{2008PhRvL.100h5003M} have shown that the rapid alignment between
 $\vec{v}$ and $\vec{B}$ is the result of a rapid relaxation process, caused by
 an interplay between pressure gradient and the magnetic field when their
 strengths are comparable.

The alignment between $\vec{J}$ and $\vec{B}$ is also critical. 
In magnetized fluids, an interesting phenomenon is the emergence of force-free fields, where the magnetic pressure much exceeds the plasma pressure, such that the Lorentz force must vanish to ensure a global balance.  This force-free field is thus a direct indication of the dominance of the magnetic energy over other energetic terms. 
 We note that the Lorentz for $\vec{F_l}$ is proportional to $\vec{J}\times\vec{B}$, and vanishes when $\vec{J}$ is parallel to $\vec{B}$. The alignment between $\vec{J}$ and $\vec{B}$ is thus a clear indication of the force-free field.

%  We compute the current $\vec{J}$ and use the angle between $\vec{B}$ and $\vec{J}$ to understand the behavior of the magnetic field. Since the Lorentz force is proportional to $\vec{J}\times\vec{B}$ and vanishes when they stay parallel,  force-free regions can be identified by measuring the angle between $\vec{B}$ and $\vec{J}$ . 

We study the alignment between $\vec{B}$, $\vec{v}$ and $\vec{J}$ at regions of different degrees of magnetization.
To quantify the importance of the magnetic field, we use the Alfven Mach number ${\cal M}_{\rm A} = \sqrt{E_{\rm k}/E_B}$, where $E_{\rm k}$ is the kinetic energy density and $E_B$ is the magnetic energy density. We study the importance of the magnetic field under different conditions as characterized by  ${\cal M}_{\rm A}$. By analyzing the alignment between the magnetic field $\vec{B}$, velocity $\vec{v}$ and current $\vec{J}$, we reveal different behaviors of the system under different ${\cal M}_{\rm A}$.

\section{Data and method}\label{enzo}
We use numerical simulations of MHD equations performed using the Enzo code \citep{2010ApJS..186..308C,2015ApJ...808...48B} with the constrained transport turned on.
The simulation conducted in this study analyzed the impact of self-gravity and magnetic fields on supersonic turbulence in isothermal molecular clouds, using high-resolution simulations and adaptive mesh refinement techniques \citep{2012ApJ...750...13C,2015ApJ...808...48B}.{\mk  They are available online \url{https://www.mhdturbulence.com/}, as presented in \cite{2020ApJ...905...14B}. 
This simulation provides a 256$^3$ cubes at super-sonic state ($M_{S,0}$ = 9) at
the resolution of 0.018 pc. The simulation we use has initial {\mk $\beta_0
\approx 0.2$}. However, as the simulation proceeds, different subregions have
different $\beta_0$ and ${\cal M}_{\rm A}$. We choose a snapshot at $t = 0.9
t_{\rm ff}$ upon
which our analyses are performed. We also investigated other snapshots and
initial conditions, and can confirm that the results, which links the behavior
of the fluid with the Alfvenic Mach number measured at the microscopic scale are
robust. The major different between those simulations reflects on the volume
filling factor of the different regimes.  }

The Alfven Mach number ${\cal M}_{\rm A}$ is the indicator of magnetic field in MHD numerical simulation:
\begin{equation}
 {\cal M}_{\rm A} = \sqrt{\frac{E_{k}}{E_B}} = \sqrt{\frac{\rho\sigma_v^2/2}{B^2/8\pi}} = \sqrt{4\pi\rho}\frac{\sigma_v}{B},
\end{equation}
which is the square root of the energy ratio between kinetic energy density $E_{\rm k}$ and magnetic energy density $E_B$.
Based on the ${\cal M}_{\rm A}$, we divide the MHD turbulence into three regimes: magnetic regime, B-k transition, and kinetic regime, and study the relation between the magnetic field and the gas motion. From the magnetic field $\vec{B}$, the  current $\vec{J}$ can be evaluated as
\begin{equation}
 J = \frac{1}{\mu_0} \nabla \times \vec{B}\;,   
\end{equation}
and we study the alignment between the magnetic field $\vec{B}$ and the current $\vec{J}$, and the alignment between the magnetic field $\vec{B}$ and the velocity $\vec{v}$ at different ${\cal M}_{\rm A}$. 
\begin{figure*}
    \centering
    \includegraphics[width = 18cm]{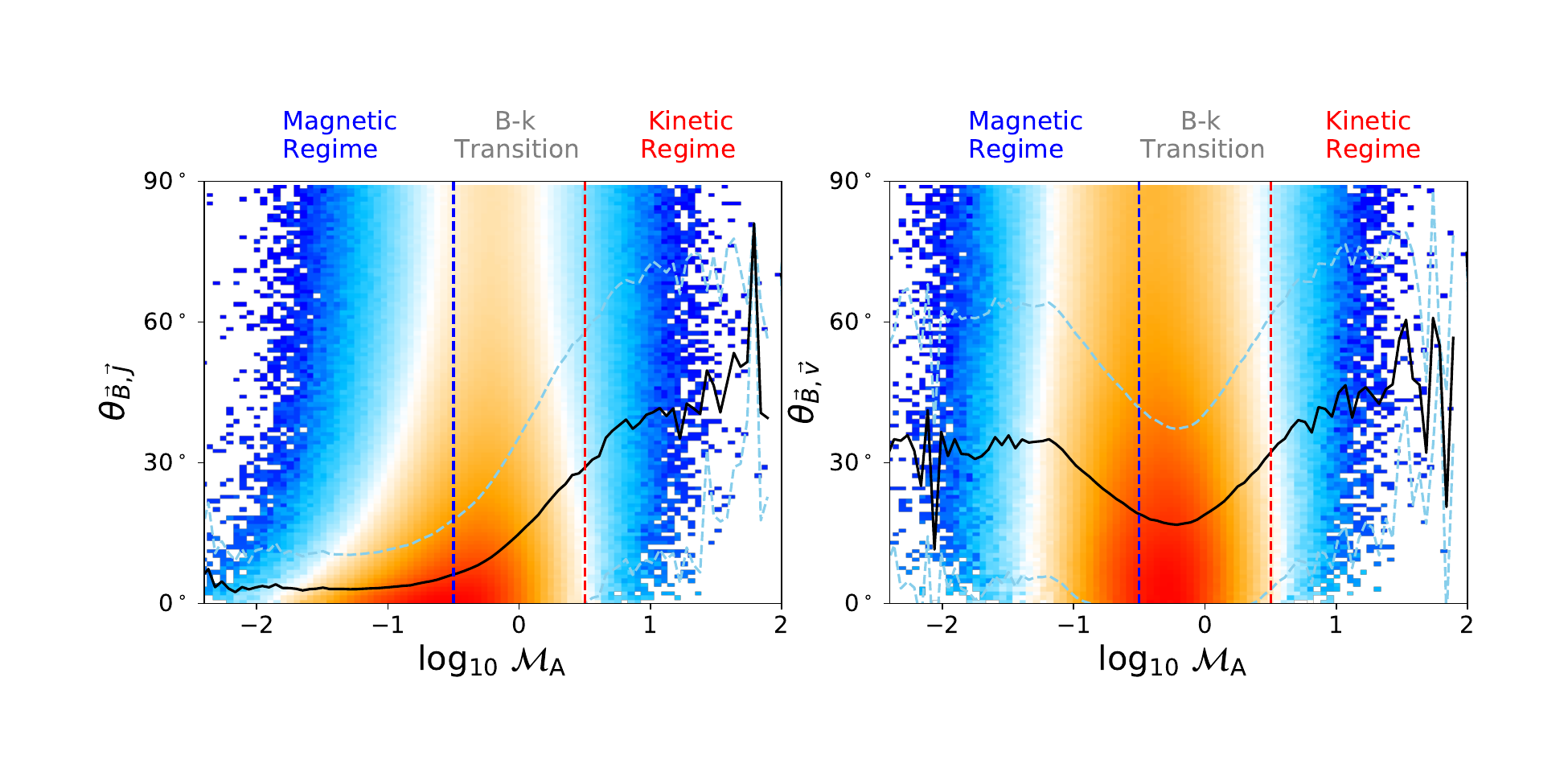}
    \caption{{\bf Distribution of ${\cal M}_{\rm A}$, $\theta_{\vec{B},\vec{J}}$, and $\theta_{\vec{B},\vec{v}}$.
    } 
    The left panel shows the distribution between Alfven Mach number ${\cal M}_{\rm A}$ and offset angle $\theta_{\vec{B}, \vec{J}}$ between the magnetic field and current.
    The right panel shows the distribution between Alfven Mach number ${\cal M}_{\rm A}$ and offset angle $\theta_{\vec{B}, \vec{v}}$ between the magnetic field and kinetic motion.
    The black line shows the main skeleton of this distribution, and the blue dash lines show its dispersion, measured at each bin.
    The blue and red dash lines show the ${\cal M}_{\rm A}\approx$ 0.3 and 3, respectively.
    }
    \label{fig1}
\end{figure*}

\begin{figure*}
    \centering
    \includegraphics[width = 18cm]{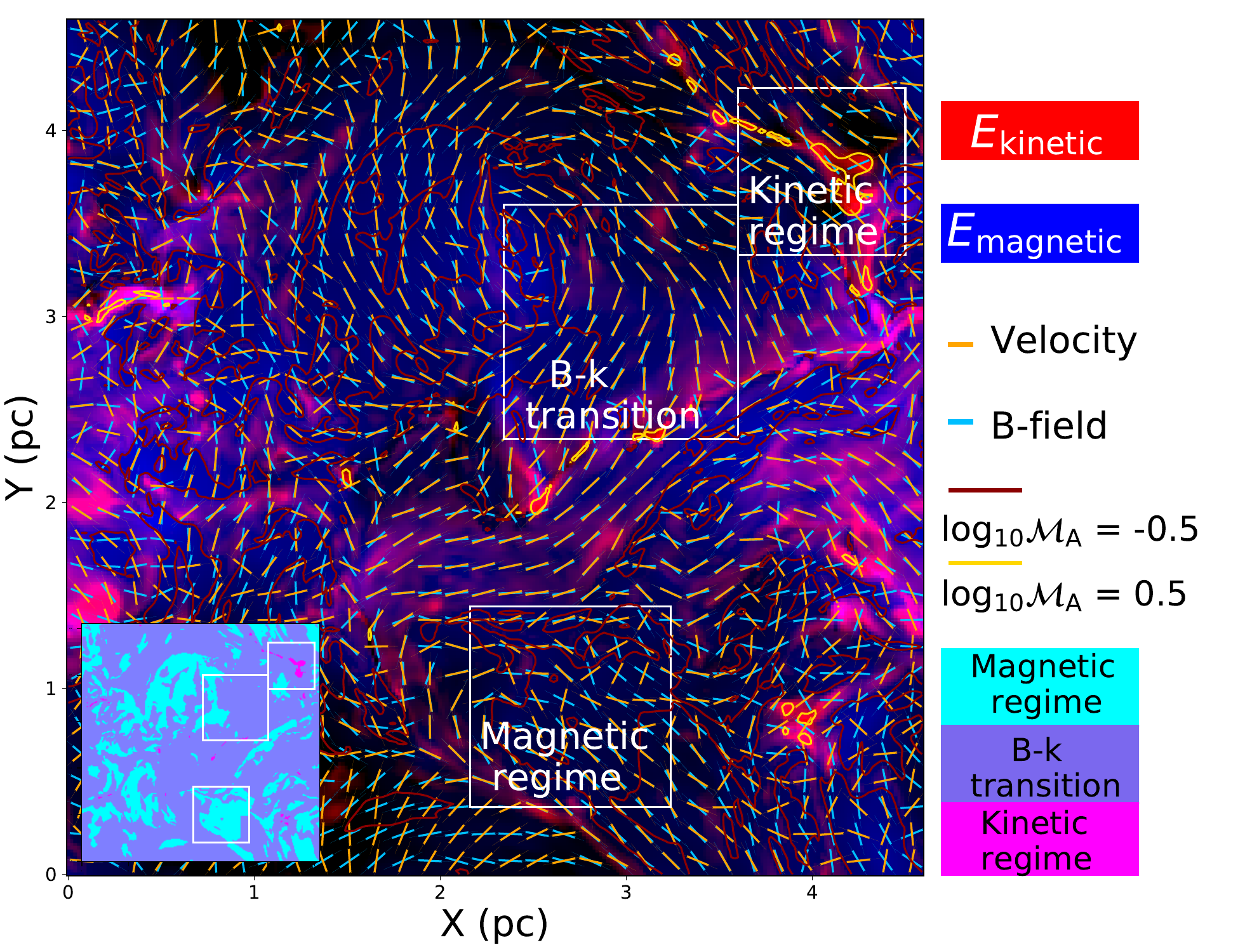}
    \caption{{\bf Alignment between magnetic field and kinetic motion.}
    The panel shows the distribution of magnetic energy density $E_B$ (blue) and kinetic energy density $E_{\rm k}$ (red).
    Orange and blue vectors present the orientation of the magnetic field and velocity.
    The contours show the -0.5 (darkred) and 0. 5 (orange) of log$_{10}$ ${\cal M}_{\rm A}$. Locations of gas belonging to different regimes are outlined in the plot at the upper left corner.  
    % The bottom panel shows the distribution of the magnetic field (blue) and current (red), while others are the same as the top panel.
    }
    \label{fig2}
\end{figure*}

\begin{figure*}
    \centering
    \includegraphics[width = 16cm]{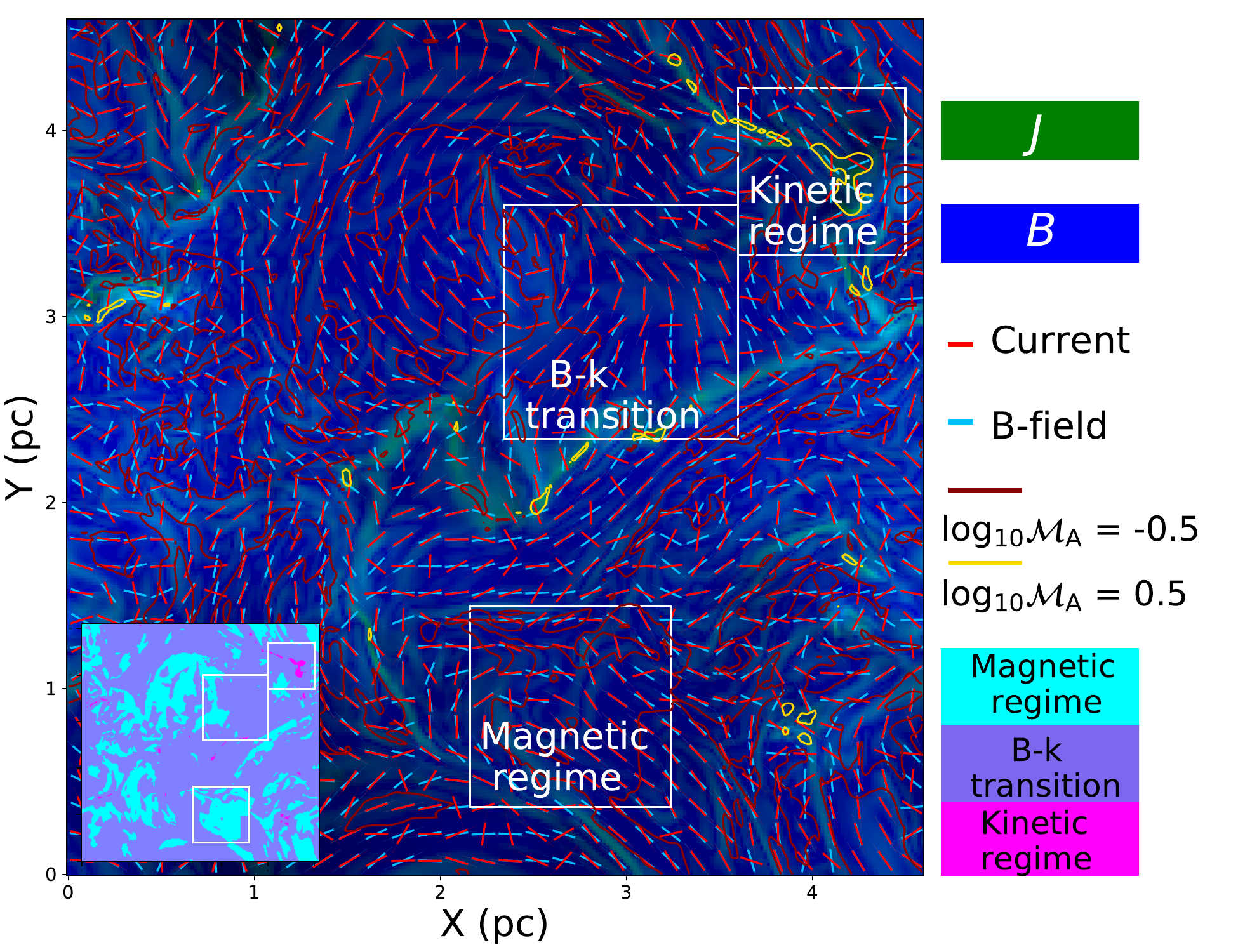}
    \caption{{\bf Alignment between magnetic field and current.}
    The panel shows the distribution of magnetic field strength $|\vec{B}|$ (blue) and current density $|\vec{J}|$ (green).
    Orange and blue vectors present the orientation of the magnetic field and velocity.
    The contours show the log$_{10}$ ${\cal M}_{\rm A}$= -0.5 (darkred) and 0.5 (orange) respectively. Locations of gas belonging to different regimes are outlined in the plot at the upper left corner.}
    \label{fig3}
\end{figure*}

\begin{figure*}
    \centering
    \includegraphics[width = 17cm]{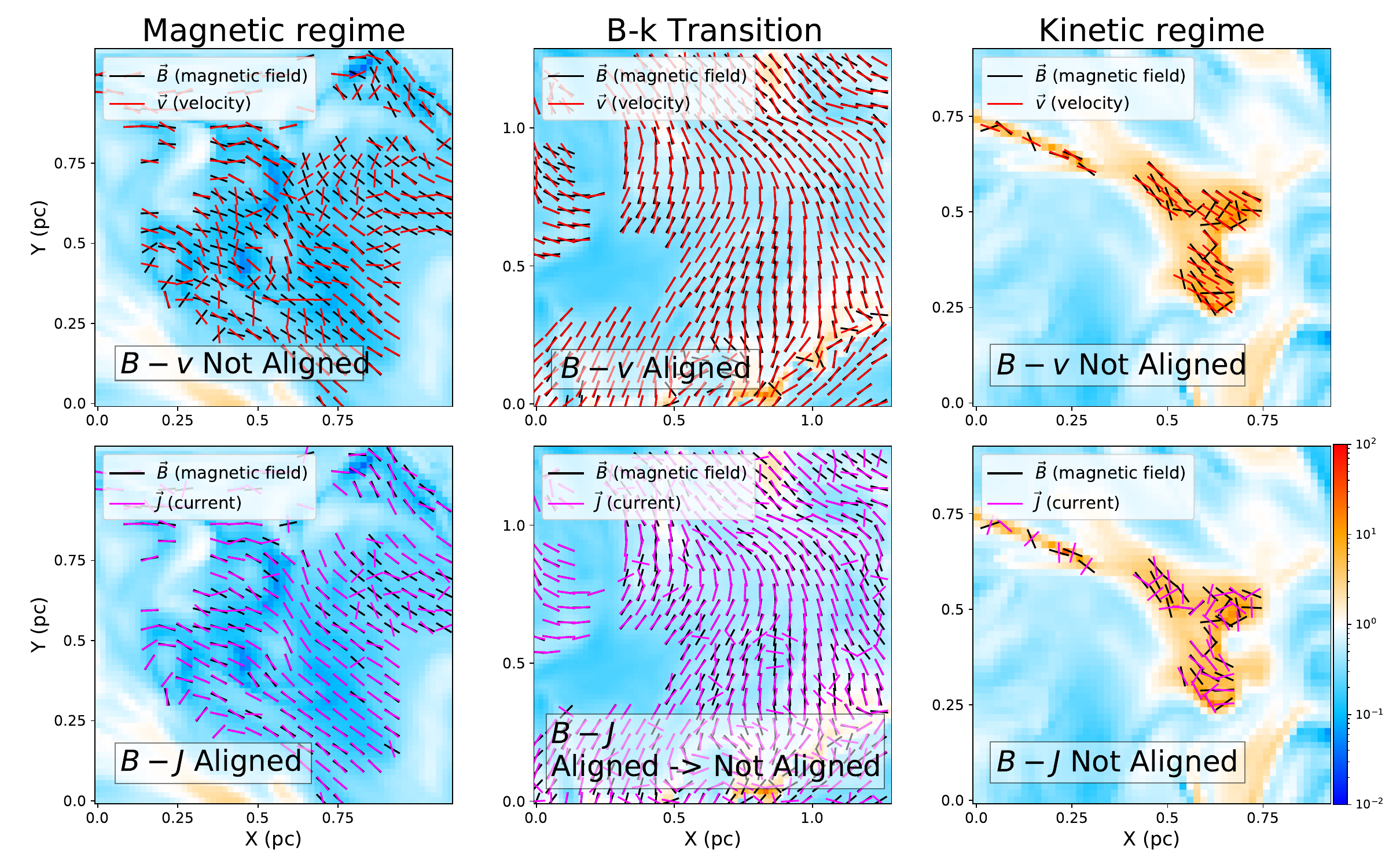}
    \caption{{\bf Alignment between magnetic field $\vec{B}$, and velocity $\vec{V}$  and current $\vec{J}$ in three different regimes.}
    The background is the ${\cal M}_{\rm A}$ distribution in the X-Y plane.
    The vectors represent the orientations of the magnetic field $\vec{B}$, velocity $\vec{v}$ and current $\vec{J}$.
    \label{fig4}}
\end{figure*}

\begin{figure*}
    \includegraphics[width=\textwidth]{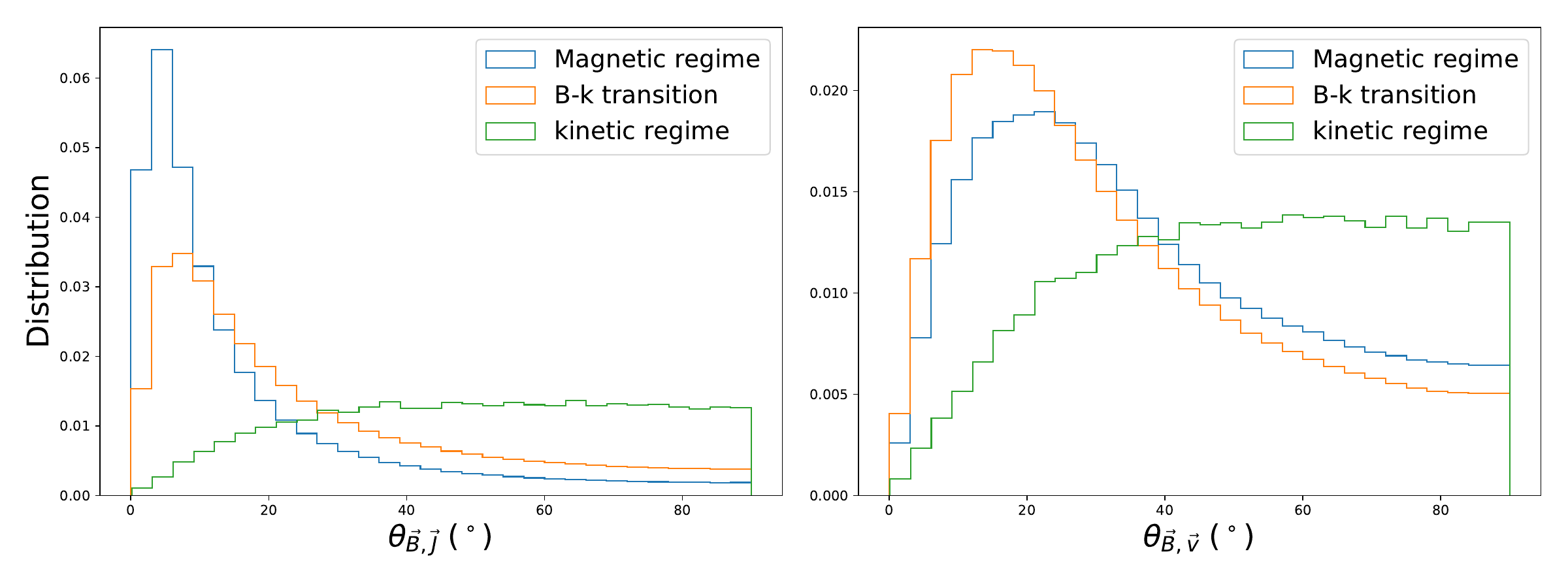}
    \caption{{\bf Distribution of the offset angle $\theta_{\vec{B}-\vec{J}}$ and $\theta_{\vec{B}-\vec{v}}$ in three regimes.\label{fig:offset}}}
\end{figure*}

\begin{figure*}
    \centering
    \includegraphics[width = 18cm]{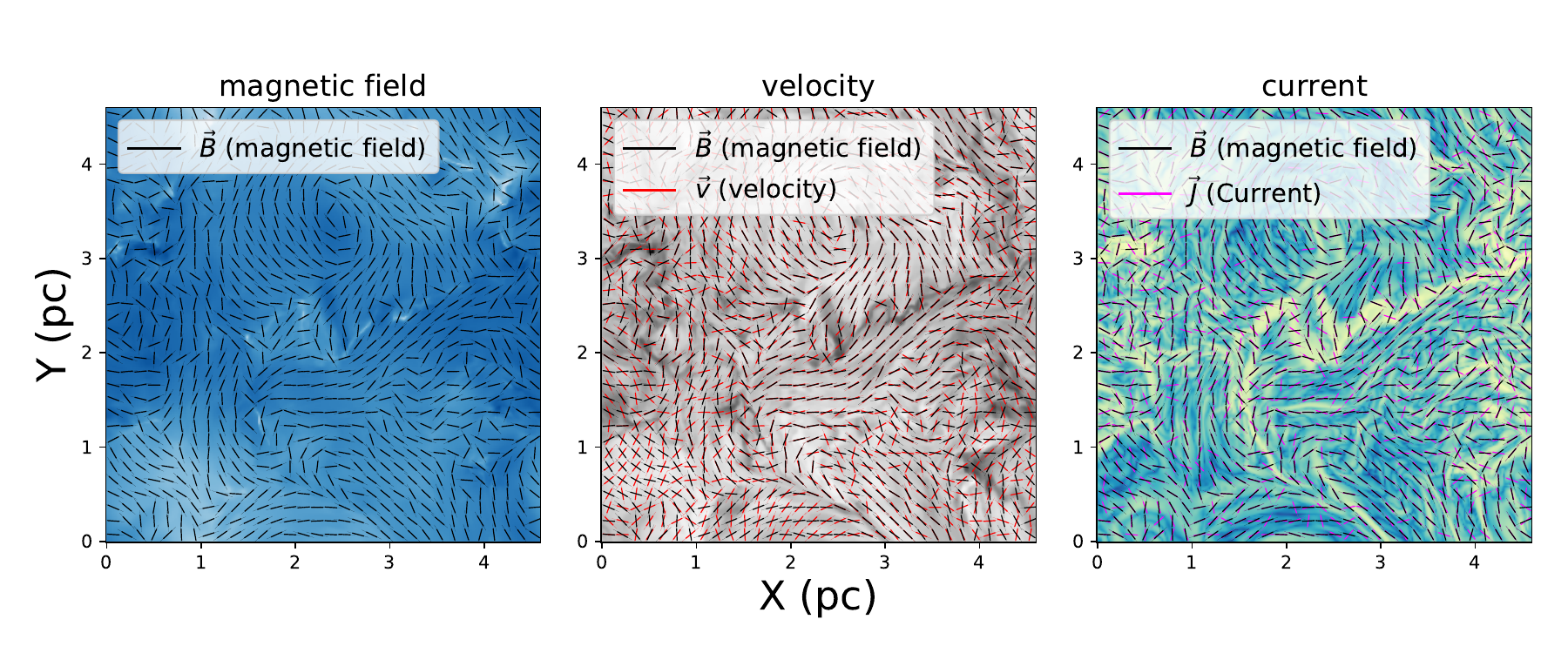}
    \caption{{\bf Distribution of magnetic field, velocity, current in the X-Y plane.}
    The background and vector in the left panel are magnetic field strength and magnetic field orientation.
    The background in the middle panel is gas density.
    The black and red vectors display the orientation of the magnetic field and velocity.
    The background in the right panel is the strength of the current.
    The purple and black vectors present the magnetic field orientation and current orientation. \label{fig:b}}
    \label{fig:enter-label}
\end{figure*}

\begin{figure*}
    \centering
    \includegraphics[width=\textwidth]{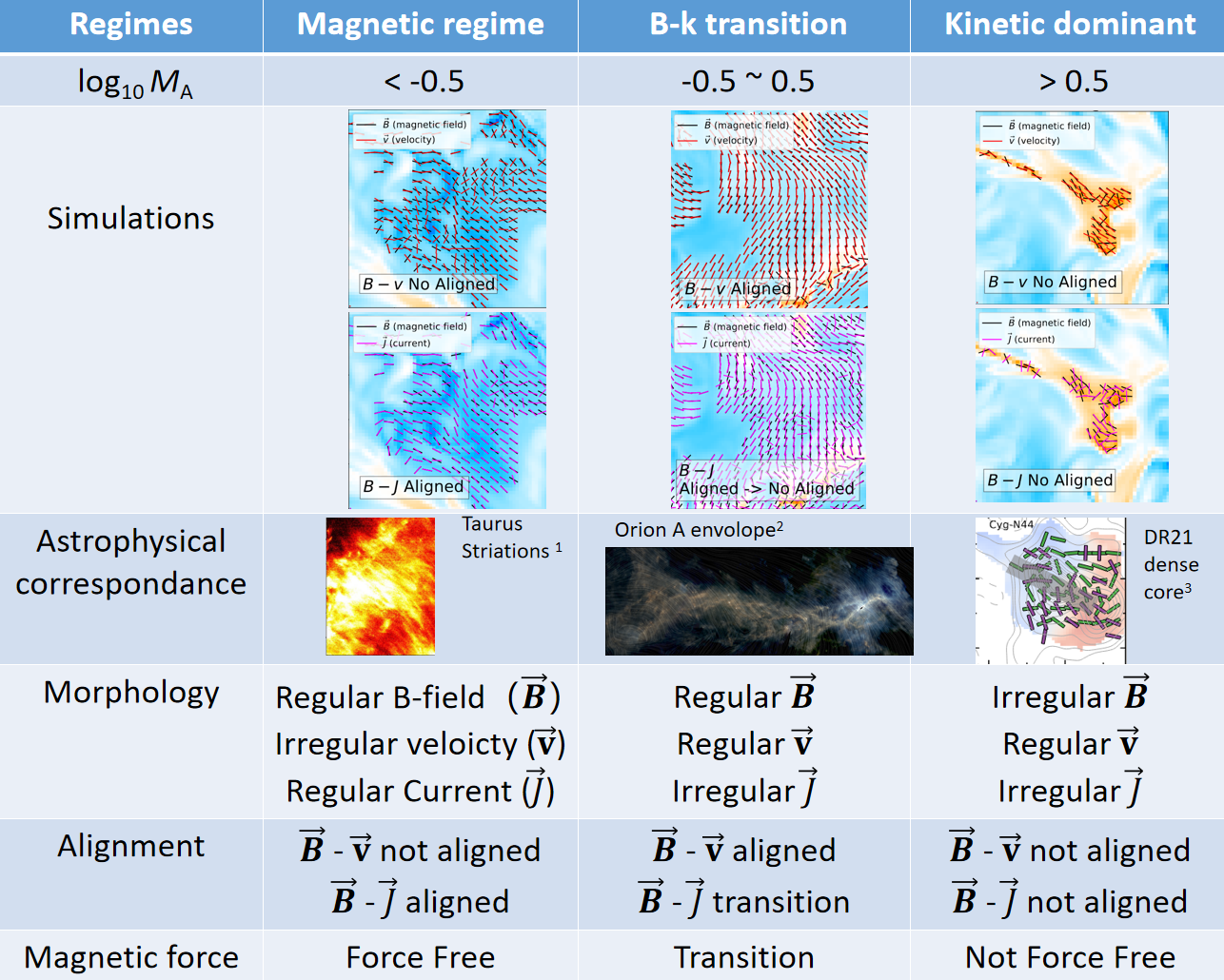}
    \caption{\bf A graphical summary of different regimes. \label{fig:regimes}}
\end{figure*}

\section{Three regimes of MHD Turbulence}

Based on the Alfven Mach number ${\cal M}_{\rm A}$, we divide the simulation box
into three regimes: magnetic regime (${\cal M}_{\rm A} < 0.3$), B-k transition
($0.3<{\cal M}_{\rm A}<3$), and kinetic regime (${\cal M}_{\rm A}>3$), and study
the alignment between the magnetic field $\vec{B}$, velocity $\vec{v}$, and
current $\vec{J}$ at different ${\cal M}_{\rm A}$. The relationship between the
magnetic field-velocity angle $\theta_{\vec{B}-\vec{v}}$, the magnetic
field-current angle $\theta_{\vec{B}-\vec{J}}$ and the Alfven Mach number is
shown in Fig.\,\ref{fig1}. When plotting these distributions, we divide our
distribution by the
distribution expected if the vectors are randomly oriented, and focus on the
additional alignment caused by \emph{physics} \footnote{The probability density function, p($\theta$), of the angle between two vectors of random distribution is distributed in an N-dimensional space as:
\begin{equation}
    p(\theta) = \frac{\Gamma(\frac{n}{2})}{\Gamma(\frac{n-1}{2})} \frac{{\rm sin}^{n-2}(\theta)}{\sqrt{\pi}}
\end{equation}
where n is numbers of dimensions, and $\theta$ is the angle between two vectors of random distribution. 
In our alignment analysis, we study the angle between qualities measured in 3D. 
When $n=3$, we have
\begin{equation}
    p(\theta) = \frac{1}{2} {\rm sin}\theta \;,
\end{equation}
which two randomly-selected vectors have a alignment angle clustered at around $\theta=45^{\circ}$, caused by the projection effect.  
To remove the projection effect, we use the probability density function p($\theta$) to weight the angle distribution: 
\begin{equation}
   I( \theta_{\rm corrected}) = I(\theta_{\rm original}) / p(\theta)
\end{equation}
where $I(\theta_{\rm original})$ represents the original angle distribution and $I(\theta_{\rm corrected})$ represents the corrected angle distribution, with the projected effects removed.
}. 

With increasing ${\cal M}_{\rm A}$, the alignment between the magnetic field $\vec{B}$ and the current $\vec{J}$ changes from aligned to not aligned, and the alignment between the magnetic field $\vec{B}$ and the velocity $\vec{v}$ change from almost no alignment ($\theta_{\vec{B}-\vec{v}} \approx 30^{\circ}$), to a weak alignment ($\theta_{\vec{B}-\vec{v}} \approx 20^{\circ}$), back to no alignment ($\theta_{\vec{B}-\vec{v}} \approx 30^{\circ}$).

We note that the magnetic force  is
\begin{equation}
 \vec{f}_{\rm L} = \vec{J} \times \vec{B}\;.
\end{equation}
When $\vec{J}$ is parallel to $\vec{B}$, the magnetic force vanishes, and this field configuration is called the force-free configuration. The magnetic force is activated if the angle between $\vec{J}$ and $\vec{B}$ is large.

The monotonic decrease of alignment between $\vec{B}$ and $\vec{J}$  at increasing   ${\cal M}_{\rm A}$  is related to the decrease in the importance of the magnetic field, leading to the system moving away from the  \emph{force-free} regime. Based on the Alfven Mach number and the behavior of the system, we divide the simulation into three regimes: \begin{itemize}
    \item The magnetic regime (${\cal M}_{\rm A}<0.3$): the magnetic energy is
    far above the local kinetic energy, where the $\Vec{B}$ and $\Vec{v}$ do not
    stay aligned, yet the $\Vec{B}$ and $\Vec{J}$ are aligned.
    \item The transition regime (B-k transition, $0.3<{\cal M}_{\rm A}<3$): the magnetic energy and kinetic energy have similar densities.  the $\Vec{B}$ and $\Vec{v}$ stay aligned, and the $\Vec{B}$ and $\Vec{J}$ evolve from aligned to not aligned as  ${\cal M}_{\rm A}$ increases.
    \item The kinetic regime: the kinetic energy is far above the local magnetic energy, the $\Vec{B}$ and $\Vec{v}$ are not aligned, and  $\Vec{B}$ and $\Vec{J}$ are not aligned.
\end{itemize}
which are plotted in Figs. \ref{fig2}, \ref{fig3} and \ref{fig4}, and are
summarized in Fig. \ref{regimes}. Some addition slices to the simulation box can
be found in Fig. \ref{fig:b}. We acknowledge that there is not clear boundary between these
regimes. However, after dividing the simulation box into these three regimes,
the alignment angle between the magnetic field $\vec{B}$, velocity $\vec{v}$,
and current $\vec{J}$ do have distinct distributions (Fig. \ref{fig:offset}). 
2D slice of magnetic field, velocity  and current can be found in
Fig.\,\ref{fig:b}.

\subsection{Magnetic Regime}\label{magnetic regime}
The first regime we discovered is the magnetic regime, where the magnetic energy is far above the local kinetic energy ($\cal{M}_{\rm A}<$ 0.3). This regime has two properties, \emph{the alignment between  $\Vec{B}$ and $\Vec{J}$ which points to the formation of a force-free field} and the \emph{lack of alignment between $\Vec{B}$ and $\Vec{v}$}. 

The strong alignment between $\Vec{B}$ and $\Vec{J}$ indicates that the field configuration is force-free, with other forces being dynamically unimportant. We also observe a lack of alignment between $\Vec{B}$ and $\Vec{v}$, which challenges the common understanding of the magnetic field being ``wires'' that guide the motion of the gas.  In contrast, the motion of gas does not appear to stay aligned with the orientation of the magnetic field line. A strong magnetic field does not necessarily lead to motions that follow the field lines.

From astronomy observations, we identify the magnetic regime from low-density regions around some existing molecular clouds. One such example is the existence of striations located at the outer part of the Taurus molecular cloud \citep{2016MNRAS.461.3918H,2016MNRAS.462.3602T}.

\subsection{B-k Transition}
At $0.3<\cal{M}_{\rm A}$ $<3$, where the magnetic energy and the kinetic energy are similar, we find a transition regime characterized by a breakdown of the alignment between $\vec{B}$ and $J$, and a strong alignment between  $\vec{B}$ and $\vec{v}$. The breakdown of the $\vec{B}$-$J$ alignment results from a decrease in the importance of the magnetic field. However, the alignment between  $\vec{B}$ and $\vec{v}$  deserves further discussions.

We find the  $\vec{B}$ and $\vec{v}$ can achieve alignment in the B-k transition regime, where the magnetic and kinetic energy have similar densities. This finding challenges the common understanding that the alignment between $\vec{B}$ and $\vec{v}$ indicates the dominance of the magnetic field. This alignment has been found by \cite{2008PhRvL.100h5003M},  which results from a ``rapid and robust relaxation process in turbulent flows''. Our findings support their conclusion. However, different from the claim that `` the alignment of the velocity and magnetic field fluctuations occurs rapidly in magnetohydrodynamics for a variety of parameters'', in our case, this alignment only occurs at the transition regime with moderate $\cal{M}_{\rm A}$. The fact that this alignment can occur only at a particular regime of of $\cal{M}_{\rm A}$ agrees with the findings of \cite{2017NJPh...19f5003K}, where our critical $\cal{M}_{\rm A}$ should correspond to their critical density. 
In the B-k transition regime, the {\mk $\vec{B}$} and $\vec{J}$ are {\mk moderately aligned} with $\theta \approx 15^{\circ}$.

% where the magnetic field does not need to dominant. This result is consistent with our finding, where the alignment between $\vec{B}$ and $\vec{v}$ only occurs at the transition regime, not in the magnetic regime as previously believed.

% often taken as an evidence that the magnetic field dominant over other forces. 

% We conclude that directional alignment is a rapid and
% robust process in turbulence. 

% At this transition regime, we observe a relatively strong alignment between $\vec{B}$ and $\vec{v}$ with $\theta_{\vec{B}-\vec{v}} \approx 20 \pm 20$, where in most cases, the velocity field appears to stay aligned with the magnetic field. What is interesting is that the alignment between $\vec{B}$ and $\vec{v}$ only occurs at this transition regime, not in the magnetic regime as previously believed. 

From the reported values of the Alfven Mach numbers (or the mass-to-flux ratio) \cite{2023ASPC..534..193P}, we believe that a large number of observations of the observations are probing gas located in this transition regime. Examples include the envelope of 
massive star formation Orion\,A \citep{2022ApJ...934...45Z}, and other star formation regions such as Taurus, L1551 and so on \citep{2019NatAs...3..776H,2021ApJ...912....2H}.

\subsection{Kinetic Regime}\label{kinetic regime}

At the kinetic regime with high kinetic energy (high ${\cal M}_{\rm A}$), the $\Vec{B}-\Vec{v}$ not aligned, and the $\Vec{B}-\Vec{J}$ not aligned. The lack of this alignment is the result of the weak magnetic field.
In astronomical observations, this corresponds to dense, collapsing regions with strong turbulence, such as the Cyg N44 in Dr21 \citep{2018ApJ...865..110C}.

% where turbulent kinetic dominates whole physical processes with irregular magnetic field, regular velocity field and irregular current field.
% The Kinetic regime of MHD turbulence occurs in high ${\cal M}_{\rm A}$, where the magnetic energy is far below kinetic energy.
% This regime is opposite to the magnetic regime (see Sect.\,\ref{magnetic regime}).
% MHD turbulence in this regime has regular velocity field structures and irregular magnetic field, current morphology.
% The kinetic motion of MHD turbulence dominates all the physical processes of the kinetic regime.
% The uniform velocity field is derived by strong kinetic motion, which could ignore the effect of the magnetic field and even distort the local magnetic field.

% The current change with magnetic field change (Ampere Law: J = $\nabla\times B/\mu_0$) and present the irregular structure, which electromagnetic field against strong extra force from turbulent motion to form a pinch with an axial current and a toroidal magnetic field \citep{1978Ap&SS..55..487A}.

\section{Conclusions}
We investigate the effect of a magnetic field on supersonic turbulence. By evaluating the Alfven Mach number $\cal{M}_{\rm A}$ at different locations in a turbulence box and studying the alignment between the magnetic field $\vec{B}$, the velocity $\vec{v}$ and the current $\vec{J}$, we reveal the different behavior of the system under various conditions. These regimes include:

\begin{itemize}

 \item The magnetic regime (${\cal M}_{\rm A}<0.3$): the magnetic energy is far above the local kinetic energy, where the $\Vec{B}-\Vec{v}$ do not stay aligned, yet the $\Vec{B}-\Vec{J}$ aligned. The magnetic field is force-free.

 \item The transition regime (B-k transition, $0.3<{\cal M}_{\rm A}<3$): the
 magnetic energy and kinetic energy have similar densities.  $\Vec{B}$ and
 $\Vec{v}$ are aligned, and the $\Vec{B}$ and $\Vec{J}$ are also aligned. We
 note this alignment between $\vec{B}$ and $\vec{v}$ does not necessarily
 imply a
 strong, dominant magnetic field but can also result from a rapid relaxation process in turbulent flows.

 \item The kinetic regime: the kinetic energy is far above the local magnetic energy, the $\Vec{B}-\Vec{v}$ not aligned, and the $\Vec{B}-\Vec{J}$ not aligned.

\end{itemize}

They are summarized in \ref{fig4}. Since there is a correlation between the
Alfven Mach number and the gas density e.g. \citep{2024ApJ...976..209Z}, the
magnetic regime exists in the lower-density part and the kinetic regime in the
higher-density part. The transition regime is an intermediate state between the
two. Using observational data, we find cases that support the existence of these
regimes.

The results guide the interpretation of new observations. It breaks down the common understanding of the magnetic field as a rigid wire that guides gas motion and replaces it with the complex behavior of the gas under different conditions. The alignment between $\vec{B}$ and $\vec{J}$ points to the dominance of the magnetic field, and the alignment between $\vec{B}$ and $\vec{v}$ is likely the result of a rapid self-organization process in turbulent flows. Some supporting cases are identified from observations of the interstellar medium. 

We reveal various regimes where the fluid behaves differently under different
conditions. To our knowledge, this is the first time these different regimes are
clearly outlined. The alignment between these quantities has been studied in a
recent paper \citep{2024arXiv240516626B} where the authors reported strong
alignments in the strongly magnetized regime and some scale-dependence alignment
behavior. We are revealing a much clearer picture with the Alfven Mach number as
the only parameter dictating the behavior of the fluid and the alignments.

% {\mk
% Recently, there has been a rapid increase in the use of Machine-Learning models to analyze these simulations, where \citet{2019ApJ...882L..12P} find that neural networks can distinguish between different regimes of MHD turbulence by taking advantage of ridge-like features, using density-slices. A detailed division of simulations into regimes, which we articulate in this paper, can be helpful in building and testing future machine-learning models.
% The link between the ridges and the alignment between $\vec{B}$,  $\vec{v}$ and $\vec{j}$ with those ridges deserves further investigation.\\}

% The
%  mechanism leading to the alignment between $\vec{B}$ and $\vec{v}$ in the transition regime remains to be understood.
% {\mk
% {\it Data:} The data we use in this work is from the MHD simulation of \cite{2012ApJ...750...13C,2015ApJ...808...48B}, which is publicly available at \url{https://www.mhdturbulence.com/} presented in \citet{2020ApJ...905...14B}.}

% ----------------------------------------------

\section*{Data availability}
The data used in this work is from the MHD simulation of \cite{2012ApJ...750...13C,2015ApJ...808...48B}, which is publicly available at \url{https://www.mhdturbulence.com/} presented in \citet{2020ApJ...905...14B}.

\section*{Acknowledgements}
% Part of this work was supported by the German
      % \emph{Deut\-sche For\-schungs\-ge\-mein\-schaft, DFG\/} project
      % number Ts~17/2--1.
      GXL acknowledges support from
NSFC grant No. 12273032 and 12033005. We thank Blakesley Burkhart for discussions on the maching-learning paper and contributors of \url{https://www.mhdturbulence.com/} for making the data publicly available.
% \end{acknowledgements}

\appendix

% % Background: Gas density, Change?

% \begin{figure}
%     \centering
%     \includegraphics[width = 16cm]{ff/EBEvrgb_00.pdf}
%     \caption{Caption}
%     \label{fig:enter-label}
% \end{figure}

% \begin{figure}
%     \centering
%     \includegraphics[width = 17cm]{ff/BJvabc_00.pdf}
%     \caption{Caption}
%     \label{fig:enter-label}
% \end{figure}

% \section{Removing the projection effect in angle distributions} \label{sec:remove}

% \begin{figure}
%     \centering
%     \includegraphics[width=10cm]{f/ptheta.png}
%     \caption{\textbf{Distribution of probability density function p($\theta$) of angle between two vectors of random distribution.}
%     Different colors represent different spatial dimension. }
%     \label{figp}
% \end{figure}


\bibliographystyle{mnras}
\bibliography{sn-bibliography}

\end{document}